# The $\alpha\alpha_s^2$ corrections to the first moment of the polarized virtual photon structure function $g_1^\gamma(x,Q^2,P^2)$

Ken Sasaki[a)*], Takahiro Ueda[a)†], Tsuneo Uematsu[b)‡]

a) Dept. of Physics, Faculty of Engineering, Yokohama National University
Yokohama 240-8501, JAPAN

b) Dept. of Physics, Graduate School of Science, Kyoto University
Yoshida, Kyoto 606-8501, JAPAN

**Abstract**

We present the next-to-next-to-leading order ($\alpha\alpha_s^2$) corrections to the first moment of the polarized virtual photon structure function $g_1^\gamma(x,Q^2,P^2)$ in the kinematical region $\Lambda^2 \ll P^2 \ll Q^2$, where $-Q^2(-P^2)$ is the mass squared of the probe (target) photon and $\Lambda$ is the QCD scale parameter. In order to evaluate the three-loop-level photon matrix element of the flavor singlet axial current, we resort to the Adler-Bardeen theorem for the axial anomaly and we calculate in effect the two-loop diagrams for the photon matrix element of the gluon operator. The $\alpha\alpha_s^2$ corrections are found to be about 3% of the sum of the leading order ($\alpha$) and the next-to-leading order ($\alpha\alpha_s$) contributions, when $Q^2 = 30 \sim 100\text{GeV}^2$ and $P^2 = 3\text{GeV}^2$, and the number of active quark flavors $n_f$ is three to five.

[*]e-mail address: sasaki@phys.ynu.ac.jp
[†]e-mail address: t-ueda@phys.ynu.ac.jp
[‡]e-mail address: uematsu@phys.h.kyoto-u.ac.jp

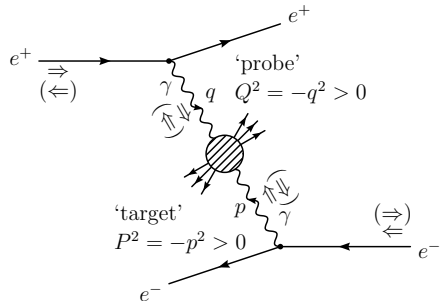

Figure 1: Deep inelastic scattering on a virtual photon in the polarized $e^+\ e^-$ collider experiments.

# 1   Introduction

The investigation of the photon structure has been an active field of research both theoretically and experimentally in recent years [1, 2, 3, 4, 5]. Also there has been growing interest in the study of the spin structure of the photon. In particular, the first moment of the polarized photon structure function $g_1^\gamma$ has attracted much attention in connection with its relevance to the QED and QCD axial anomaly [6, 7, 8, 9, 10]. The polarized photon structure functions can be measured from two-photon processes in the polarized $e^+e^-$ collider experiments as shown in Fig. 1, where $-Q^2(-P^2)$ is the mass squared of the probe (target) photon. The photon structure functions are defined in the lowest order of the QED coupling constant $\alpha = e^2/4\pi$ and, in this paper, they are of order $\alpha$.

For a real photon ($P^2 = 0$) target, there exists only one spin-dependent structure function $g_1^\gamma(x, Q^2)$. The QCD analysis of $g_1^\gamma$ was performed in the leading order (LO) (the order $\alpha$) [11], and in the next-to-leading order (NLO) (the order $\alpha\alpha_s$) [12, 13], where $\alpha_s = g^2/4\pi$ is the QCD coupling constant. In the case of a virtual photon target ($P^2 \neq 0$) there appear two spin-dependent structure functions, $g_1^\gamma(x, Q^2, P^2)$ and $g_2^\gamma(x, Q^2, P^2)$. The former has been investigated up to the NLO by the present authors in [14, 15], and also in the second paper of [13]. In Refs.[14, 15] the structure function $g_1^\gamma(x, Q^2, P^2)$ was analyzed in the kinematical region

$$\Lambda^2 \ll P^2 \ll Q^2\ , \tag{1.1}$$

where $\Lambda$ is the QCD scale parameter. The advantage in studying the virtual photon target in the kinematical region (1.1) is that we can calculate structure functions by the perturbative method without any experimental data input [16], which is contrasted with the case of a real photon target where in the NLO there exist nonperturbative pieces [17].

In the present paper we focus on the photon structure function $g_1^\gamma$, and especially on the first moment sum rule of $g_1^\gamma(x, Q^2, P^2)$ in the kinematical region (1.1). We present a new result on the next-to-next-to-leading order (NNLO) (the order $\alpha\alpha_s^2$) corrections to this sum rule.

The polarized structure function $g_1^\gamma$ of the real photon satisfies a remarkable sum rule [6, 7, 8, 9, 10]

$$\int_0^1 g_1^\gamma(x, Q^2) dx = 0 \ . \tag{1.2}$$

In particular, applying the Drell-Hearn-Gerasimov sum rule [18] to the case of a virtual photon target and using the fact that the photon has zero anomalous magnetic moment, the authors of Ref. [10] showed that the sum rule, (1.2), holds to all orders in perturbation theory in both QED and QCD.

When the target photon becomes off-shell, i.e., $P^2 \neq 0$, the first moment of the corresponding photon structure function $g_1^\gamma(x, Q^2, P^2)$ does not vanish any more. In fact, for the case $\Lambda^2 \ll P^2 \ll Q^2$, the first moment has been calculated up to the NLO ($\alpha\alpha_s$), as follows [8, 14];

$$\begin{aligned}
\int_0^1 dx g_1^\gamma(x, Q^2, P^2) &= -\frac{3\alpha}{\pi} \left[ \sum_{i=1}^{n_f} e_i^4 \left( 1 - \frac{\alpha_s(Q^2)}{\pi} \right) \right. \\
&\quad \left. - \frac{2}{\beta_0} (\sum_{i=1}^{n_f} e_i^2)^2 \left( \frac{\alpha_s(P^2)}{\pi} - \frac{\alpha_s(Q^2)}{\pi} \right) \right] + \mathcal{O}(\alpha\alpha_s^2), \quad (1.3)
\end{aligned}$$

with $\beta_0 = 11 - 2n_f/3$ being the one-loop QCD $\beta$ function. Here $\alpha_s(Q^2)$ is the QCD running coupling constant, and $e_i$ is the electromagnetic charge of the active quark (i.e., the massless quark) with flavor $i$ in the unit of proton charge and $n_f$ is the number of active quark flavors. The first moment depends only on the average charge squared $\langle e^2 \rangle = (\sum_i^{n_f} e_i^2)/n_f$ and the average of the fourth power of the charge $\langle e^4 \rangle = (\sum_i^{n_f} e_i^4)/n_f$. Note that the r.h.s. of Eq.(1.3) does not involve any experimental data input. This is because the nonperturbative pieces, which take part in the real photon target case, can be neglected in the kinematical region (1.1) and the whole twist-two contributions to $g_1^\gamma(x, Q^2, P^2)$ can be computed by the perturbative method. It is also noted that the first term in the square brackets of the r.h.s. of Eq.(1.3) resulted from the QED triangle anomaly while the second term comes from the QCD triangle anomaly [8, 15].

## 2 Theoretical framework based on OPE

For the analysis of the NNLO ($\alpha\alpha_s^2$) corrections to the first moment of $g_1^\gamma(x, Q^2, P^2)$, we will apply the framework of the operator product expansion (OPE) supplemented by the renormalization group method. First recall that for the OPE of two electromagnetic (and thus gauge-invariant) currents, only gauge-invariant operators need to be included with their renormalization basis [19]. Since gauge-invariant twist-two gluon and photon operators with spin one are absent, we consider only quark operators, i.e., the flavor singlet $J_{5S}^\sigma$ and nonsinglet $J_{5NS}^\sigma$ axial currents, as follows:

$$J_{5S}^\sigma = \overline{\psi}\gamma^\sigma\gamma_5\, 1\psi\ , \qquad J_{5NS}^\sigma = \overline{\psi}\gamma^\sigma\gamma_5(Q_{ch}^2 - \langle e^2\rangle 1)\psi\ , \qquad (2.1)$$

where 1 is an $n_f \times n_f$ unit matrix and $Q_{ch}^2$ is the square of the $n_f \times n_f$ quark-charge matrix so that $\text{Tr}(Q_{ch}^2 - \langle e^2\rangle 1) = 0$. Writing the photon matrix elements of the quark currents as

$$\langle\gamma(p)|J_{5i}^\sigma(\mu^2)|\gamma(p)\rangle = \epsilon_\lambda^*[-2i\epsilon^{\sigma\lambda\tau\rho}p_\rho]\epsilon_\tau\, \langle\gamma(p)||J_{5i}(\mu^2)||\gamma(p)\rangle\ , \qquad i = S, NS \quad (2.2)$$

where $\epsilon_\mu^*$ and $\epsilon_\tau$ are the polarization vectors of the target photon with momentum $p$ and $\mu$ is the renormalization point, then the first moment sum rule of $g_1^\gamma(x, Q^2, P^2)$ is expressed as

$$\begin{aligned}\int_0^1 dx g_1^\gamma(x, Q^2, P^2) &= \langle\gamma(p)||J_{5S}(\mu^2)||\gamma(p)\rangle\, C_S(Q^2/\mu^2, \bar{g}(\mu^2), \alpha) \\ &+ \langle\gamma(p)||J_{5NS}(\mu^2)||\gamma(p)\rangle\, C_{NS}(Q^2/\mu^2, \bar{g}(\mu^2), \alpha)\ . \quad (2.3)\end{aligned}$$

Here $C_S$ and $C_{NS}$ are the coefficient functions corresponding to the currents $J_{5S}^\sigma$ and $J_{5NS}^\sigma$, respectively. Putting it more closely, $C_S$ and $C_{NS}$ are the $n=1$ coefficient functions which appear in the OPE of two electromagnetic currents. Throughout this paper we neglect the effect of quark masses.

We choose the renormalization point at $\mu^2 = P^2$. For $-p^2 = P^2 \gg \Lambda^2$, we can calculate perturbatively the photon matrix elements of the axial currents, which are expressed in the form as

$$\langle \gamma(p) || J_{5i}(\mu^2 = P^2) || \gamma(p) \rangle = \frac{\alpha}{4\pi} A_i , \qquad i = S, NS , \qquad (2.4)$$

with

$$A_i = A_i^{(0)} + \frac{\alpha_s(P^2)}{4\pi} A_i^{(1)} + \left(\frac{\alpha_s(P^2)}{4\pi}\right)^2 A_i^{(2)} + \cdots . \qquad (2.5)$$

The leading terms $A_S^{(0)}$ and $A_{NS}^{(0)}$ are connected with the Adler-Bell-Jackiw anomaly [20] and are already known [21]. In Sec. 4 we will show, using the nonrenormalization theorem for the triangle anomaly [22],

$$A_S^{(1)} = A_{NS}^{(1)} = A_{NS}^{(2)} = 0 . \qquad (2.6)$$

In fact, the result $A_S^{(1)} = A_{NS}^{(1)} = 0$ has been used in Ref.[14] to obtain the NLO ($\alpha\alpha_s$) corrections to the sum rule shown in Eq.(1.3). On the other hand, $A_S^{(2)}$ is nonvanishing and will be calculated in Sec. 4.

The $Q^2$ dependence of the coefficient functions $C_S$ and $C_{NS}$ is governed by the renormalization group equations. The solutions to these equations are given by

$$C_i(Q^2/P^2, \bar{g}(P^2), \alpha) = \exp\left[\int_{\bar{g}(Q^2)}^{\bar{g}(P^2)} dg' \frac{\gamma_i(g')}{\beta(g')}\right] C_i(1, \bar{g}(Q^2), \alpha) , \qquad i = S, NS \qquad (2.7)$$

where $\gamma_i(g)$ is the anomalous dimension of the axial current $J_{5i}^\sigma$ and $\beta(g)$ is the QCD $\beta$-function. We expand $\gamma_i(g)$ in powers of $g$ as

$$\gamma_i(g) = \gamma_i^{(0)} \frac{g^2}{16\pi^2} + \gamma_i^{(1)} \left(\frac{g^2}{16\pi^2}\right)^2 + \gamma_i^{(2)} \left(\frac{g^2}{16\pi^2}\right)^3 + \mathcal{O}(g^8) , \qquad i = S, NS . \qquad (2.8)$$

Since the flavor nonsinglet axial current $J_{5NS}^\sigma$ is conserved in the massless limit, it undergoes no renormalization, and we have

$$\gamma_{NS}^{(0)} = \gamma_{NS}^{(1)} = \gamma_{NS}^{(2)} = \cdots = 0 . \qquad (2.9)$$

Thus the nonsinglet coefficient function $C_{NS}$ is expressed as

$$C_{NS}(Q^2/P^2, \bar{g}(P^2), \alpha) = C_{NS}(1, \bar{g}(Q^2), \alpha) . \qquad (2.10)$$

On the other hand, the flavor singlet axial current $J_{5S}^\sigma$ has a non-vanishing anomalous dimension $\gamma_S(g)$ due to the axial anomaly. To be precise, we know $\gamma_S^{(0)} = 0$ at one-loop, but at higher loops we have nonzero $\gamma_S^{(1)}$, $\gamma_S^{(2)}$ and so on. The $\beta$ function is expanded as:

$$\mu\frac{\partial g}{\partial \mu} = \beta(g) = -\beta_0 \frac{g^3}{16\pi^2} - \beta_1 \frac{g^5}{(16\pi^2)^2} - \beta_2 \frac{g^7}{(16\pi^2)^3} + \cdots . \quad (2.11)$$

Then using Eqs.(2.8) and (2.11), we obtain up to the order of $\alpha_s^2$,

$$\begin{aligned}
\exp\left[\int_{\bar{g}(Q^2)}^{\bar{g}(P^2)} dg' \frac{\gamma_S(g')}{\beta(g')}\right] &= 1 + \frac{\gamma_S^{(1)}}{2\beta_0}\left(\frac{\alpha_s(Q^2)}{4\pi} - \frac{\alpha_s(P^2)}{4\pi}\right) \\
&+ \frac{1}{4\beta_0}\left(\gamma_S^{(2)} - \gamma_S^{(1)}\frac{\beta_1}{\beta_0}\right)\left(\left(\frac{\alpha_s(Q^2)}{4\pi}\right)^2 - \left(\frac{\alpha_s(P^2)}{4\pi}\right)^2\right) \\
&+ \frac{1}{8}\left(\frac{\gamma_S^{(1)}}{\beta_0}\right)^2\left(\frac{\alpha_s(Q^2)}{4\pi} - \frac{\alpha_s(P^2)}{4\pi}\right)^2 + \mathcal{O}(\alpha_s^3) . \quad (2.12)
\end{aligned}$$

We need the information on the $\beta$ function only up to the two-loop level, i.e., $\beta_0$ and $\beta_1$, in the above expression. But later for numerical analysis, we will use the QCD running coupling constant $\alpha_s(Q^2)$ where the three-loop $\beta_2$ is also taken care of [23].

Finally the flavor singlet and nonsinglet quark coefficient functions, $C_S(1, \bar{g}(Q^2), \alpha)$ and $C_{NS}(1, \bar{g}(Q^2), \alpha)$, are expanded in power of $\alpha_s(Q^2)$ up to the two-loop level as,

$$C_S(1, \bar{g}(Q^2), \alpha) = \langle e^2\rangle\left\{1 + B_S^{(1)}\frac{\alpha_s(Q^2)}{4\pi} + B_S^{(2)}\left(\frac{\alpha_s(Q^2)}{4\pi}\right)^2 + \cdots\right\} , \quad (2.13)$$

$$C_{NS}(1, \bar{g}(Q^2), \alpha) = \left\{1 + B_{NS}^{(1)}\frac{\alpha_s(Q^2)}{4\pi} + B_{NS}^{(2)}\left(\frac{\alpha_s(Q^2)}{4\pi}\right)^2 + \cdots\right\}. \quad (2.14)$$

Then putting Eqs.(2.4-2.6), (2.10), (2.12-2.14) into Eq.(2.3), we obtain the expression for the first moment sum rule of $g_1^\gamma(x, Q^2, P^2)$ up to the NNLO $(\alpha\alpha_s^2)$ corrections as follows:

$$\begin{aligned}
\int_0^1 dx\, g_1^\gamma(x, Q^2, P^2) / \left(\frac{\alpha}{4\pi}\right) \\
= \langle e^2\rangle A_S^{(0)} + A_{NS}^{(0)} + \left(\langle e^2\rangle A_S^{(0)} B_S^{(1)} + A_{NS}^{(0)} B_{NS}^{(1)}\right)\frac{\alpha_s(Q^2)}{4\pi} \\
+ \langle e^2\rangle A_S^{(0)} \frac{\gamma_S^{(1)}}{2\beta_0}\left[\frac{\alpha_s(Q^2)}{4\pi} - \frac{\alpha_s(P^2)}{4\pi}\right]
\end{aligned}$$

$$+\Big(\langle e^2\rangle A_S^{(0)} B_S^{(2)} + A_{NS}^{(0)} B_{NS}^{(2)}\Big)\Big(\frac{\alpha_s(Q^2)}{4\pi}\Big)^2$$

$$+\langle e^2\rangle A_S^{(0)} B_S^{(1)} \frac{\gamma_S^{(1)}}{2\beta_0} \frac{\alpha_s(Q^2)}{4\pi} \Big[\frac{\alpha_s(Q^2)}{4\pi} - \frac{\alpha_s(P^2)}{4\pi}\Big]$$

$$+\langle e^2\rangle A_S^{(0)} \frac{1}{4\beta_0}\Big(\gamma_S^{(2)} - \gamma_S^{(1)}\frac{\beta_1}{\beta_0}\Big)\Big[\Big(\frac{\alpha_s(Q^2)}{4\pi}\Big)^2 - \Big(\frac{\alpha_s(P^2)}{4\pi}\Big)^2\Big]$$

$$+\langle e^2\rangle A_S^{(0)} \frac{1}{8}\Big(\frac{\gamma_S^{(1)}}{\beta_0}\Big)^2 \Big[\frac{\alpha_s(Q^2)}{4\pi} - \frac{\alpha_s(P^2)}{4\pi}\Big]^2$$

$$+\langle e^2\rangle A_S^{(2)}\Big(\frac{\alpha_s(P^2)}{4\pi}\Big)^2 . \qquad (2.15)$$

## 3  Parameters in the $\overline{\text{MS}}$ scheme

All the quantities necessary to evaluate the NNLO ($\alpha\alpha_s^2$) corrections to the first moment of $g_1^\gamma(x, Q^2, P^2)$ have been calculated in the literature and already known, except for the photon matrix element of the flavor singlet axial current at three loops. If not otherwise mentioned, all the expressions listed in this section are the ones calculated in the modified minimal subtraction ($\overline{\text{MS}}$) scheme [24].

The flavor singlet axial current $J_{5S}^\sigma$ has an anomalous dimension starting at the two-loop order due to the axial anomaly. The two-loop [25] and three-loop [26] results are:

$$\gamma_S^{(1)} = 12 C_F n_f, \qquad (3.1)$$

$$\gamma_S^{(2)} = \Big(\frac{284}{3} C_F C_A - 36 C_F^2\Big) n_f - \frac{8}{3} C_F n_f^2, \qquad (3.2)$$

with $C_F = \frac{4}{3}$ and $C_A = 3$ in QCD. In fact, $\gamma_S^{(1)}$ was first calculated in the Pauli-Villars regularization [27]. The result $\gamma_S^{(2)}$ was obtained in the $\overline{\text{MS}}$ scheme.

It is well known that a suitable prescription is required for the renormalization of the flavor singlet and nonsinglet axial currents in the framework of dimensional regularization. For the flavor singlet case, i.e., in order to obtain $\gamma_S^{(2)}$ in Ref.[26], the

$\gamma_5$-matrix was defined as [28]

$$\gamma_\mu \gamma_5 = \frac{i}{3!} \epsilon_{\mu\rho\sigma\tau} \gamma^\rho \gamma^\sigma \gamma^\tau. \tag{3.3}$$

In addition, after the standard ultraviolet renormalization of the axial current $J_{5S}^\sigma$ in the $\overline{\text{MS}}$ scheme, an extra finite renormalization was introduced so that the one-loop character of the operator relation of the axial anomaly in QCD [22],

$$\partial_\kappa J_{5S}^\kappa = \frac{g^2}{16\pi^2} \frac{n_f}{2} \epsilon^{\mu\nu\rho\kappa} G_{\mu\nu}^a G_{\rho\kappa}^a, \tag{3.4}$$

where $G_{\mu\nu}^a = \partial_\mu A_\nu^a - \partial_\nu A_\mu^a + g f^{abc} A_\mu^b A_\nu^c$ is the gluonic field-strength tensor, is retained intact to all orders in perturbation theory.

As for the prescription for the flavor nonsinglet case [29], the $\gamma_5$-matrix defined in Eq.(3.3) is used but this violates the axial Ward identity, which is to be restored by an additional finite renormalization for $J_{5NS}^\sigma$. Then we have the nullified anomalous dimension for the nonsinglet axial current (see Eq.(2.9)).

The flavor singlet coefficient function $C_S(1, \bar{g}(Q^2), \alpha)$ was calculated up to the two-loop level [30, 31] while the nonsinglet coefficient function $C_{NS}(1, \bar{g}(Q^2), \alpha)$ up to the three-loop level [29]. The one-loop [32] and two-loop [30, 31, 33] results are

$$B_S^{(1)} = B_{NS}^{(1)} = -3C_F, \tag{3.5}$$

$$B_S^{(2)} = C_F \Big[\frac{21}{2} C_F - 23 C_A + (8\zeta_3 + \frac{13}{3}) n_f\Big], \tag{3.6}$$

$$B_{NS}^{(2)} = C_F \Big[\frac{21}{2} C_F - 23 C_A + 4 n_f\Big], \tag{3.7}$$

where $\zeta_3$ is the Riemann zeta-function ($\zeta_3 = 1.202056903\cdots$). Both $B_S^{(2)}$ and $B_{NS}^{(2)}$ were calculated in the $\overline{\text{MS}}$ scheme with the prescription for the renormalization of the flavor singlet and nonsinglet axial currents explained above.

The $\beta$ function has been calculated up to the four-loop level in the $\overline{\text{MS}}$ scheme [34, 35]. For the purpose of this paper we need the $\beta$ function up to the three-loop level (see Eq.(2.11)) and, in addition to $\beta_0$, we get

$$\beta_1 = 102 - \frac{38}{3} n_f, \tag{3.8}$$

$$\beta_2 = \frac{2857}{2} - \frac{5033}{18} n_f + \frac{325}{54} n_f^2. \tag{3.9}$$

# 4 The Adler-Bardeen theorem and the photon matrix element of the axial current

In this section we show how to calculate the photon matrix elements of the axial currents (Eqs.(2.4-2.5)) up to the three-loop level ($\alpha\alpha_s^2$). We make a full use of the Adler-Bardeen (AB) theorem [22] for the axial anomaly both in QED and in QCD. With its use, the loop-level of the Feynman diagrams to be evaluated decreases by one and the calculation becomes much simpler. Also the use of the AB theorem is legitimate from the viewpoint of the renormalization scheme (RS) dependence. We already know that we should use the same RS to compute higher-order coefficient functions and anomalous dimensions of operators in order to get the RS independent predictions. In the present case, the same RS should be employed also for the calculation of the higher-order photon matrix elements of the axial currents. Recall that the definition of the $\gamma_5$-matrix in Eq.(3.3) within the $\overline{\text{MS}}$ scheme leads to an additional finite renormalization required for the axial currents. In particular for the flavor singlet axial current, the finite renormalization is fixed so that the AB theorem for the axial anomaly may be secured. Using this RS or renormalization prescription the three-loop anomalous dimension $\gamma_S^{(2)}$ of Eq.(3.2) and the two-loop coefficient function $B_S^{(2)}$ of Eq.(3.6) have been calculated. Therefore it is a consistent procedure to apply the AB theorem for the axial anomaly to the calculation of the higher-order photon matrix elements of the axial currents.

## 4.1 The AB theorem for QED

The leading terms $A_S^{(0)}$ and $A_{NS}^{(0)}$ in Eq.(2.5) are already known [21, 14]:

$$A_S^{(0)} = -12n_f\langle e^2\rangle \,, \qquad A_{NS}^{(0)} = -12n_f(\langle e^4\rangle - \langle e^2\rangle^2) \,. \tag{4.1}$$

We will rederive the above results as an example of how to use the AB theorem for the calculation of the photon matrix elements of the axial currents.

The AB theorem for the axial current $J_5^\kappa = \overline{\psi}\gamma^\kappa\gamma_5\psi$ in QED gives

$$\partial_\kappa J_5^\kappa = \frac{e^2}{16\pi^2}\epsilon^{\mu\nu\rho\kappa}F_{\mu\nu}F_{\rho\kappa} \,, \tag{4.2}$$

where $F_{\mu\nu} = \partial_\mu A_\nu - \partial_\nu A_\mu$ is the field-strength tensor of photon. The term on the r.h.s. is the Adler-Bell-Jackiw anomaly [20]. Note that this is an operator form and holds to all orders in perturbation theory. Let us define

$$(p_1 + p_2)_\kappa R^{\kappa\lambda\tau}(p_1, p_2) \equiv i \int d^4x_1 d^4x_2 e^{i\{(p_1+p_2)x - p_1 x_1 - p_2 x_2\}}$$
$$\times \langle 0 | T \Big[ \partial_\kappa J_5^\kappa(x) A^\lambda(x_1) A^\tau(x_2) \Big] | 0 \rangle_{\text{Amputated}} \quad (4.3)$$

where $A^\lambda$ is a photon field and $T$ is the covariant time-ordered product. Actually, the expression of the r.h.s. does not depend on $x$. According to the AB theorem, this is equal to

$$i \int d^4x_1 d^4x_2 e^{i\{(p_1+p_2)x - p_1 x_1 - p_2 x_2\}}$$
$$\times \frac{e^2}{16\pi^2} \langle 0 | T \Big[ 4\epsilon^{\mu\nu\rho\kappa} \partial_\mu A_\nu(x) \partial_\rho A_\kappa(x) A^\lambda(x_1) A^\tau(x_2) \Big] | 0 \rangle_{\text{Amputated}} \, . \quad (4.4)$$

Now differentiating both sides of Eq.(4.3) with respect to $p_{1\sigma}$, and setting $p_1 = -p_2 = -p$ [36], we obtain

$$R^{\sigma\lambda\tau}(-p, p) = \int d^4x_1 d^4x_2 \, e^{ip(x_1 - x_2)}$$
$$\times \langle 0 | T \Big[ J_5^\sigma(x) A^\lambda(x_1) A^\tau(x_2) \Big] | 0 \rangle_{\text{Amputated}} \, , \quad (4.5)$$

where we have used an identity $i(x - x_1)^\sigma \partial_\kappa J_5^\kappa(x) = -i J_5^\sigma(x)$, since

$$\int d^4x \Big[ i(x - x_1)^\sigma \partial_\kappa J_5^\kappa(x) \Big] = [\text{surface terms}] - i \int d^4x \Big[ \partial_{x,\kappa}(x - x_1)^\sigma \Big] J_5^\kappa(x)$$
$$= -i \int d^4x J_5^\sigma(x) \, . \quad (4.6)$$

Note that the r.h.s. of Eq.(4.5) is nothing but the photon matrix element of $J_5^\sigma$ at zero momentum transfer. Differentiating Eq.(4.4) with respect to $p_1^\sigma$, and setting $p_1 = -p_2 = -p$, we find that $R^{\sigma\lambda\tau}(-p, p)$ is also expressed as

$$R^{\sigma\lambda\tau}(-p, p) = \frac{e^2}{16\pi^2} 4 \int d^4x_1 d^4x_2 \, e^{ip(x_1 - x_2)}$$
$$\times \epsilon^{\sigma\nu\rho\kappa} \langle 0 | T \Big[ A_\nu(x) \partial_\rho A_\kappa(x) A^\lambda(x_1) A^\tau(x_2) \Big] | 0 \rangle_{\text{Amputated}} \, . \quad (4.7)$$

The leading terms $A_S^{(0)}$ and $A_{NS}^{(0)}$ are obtained from the evaluation of the r.h.s. of Eq.(4.5) in the leading order $e^2$. Instead we compute the r.h.s. of Eq.(4.7). It is

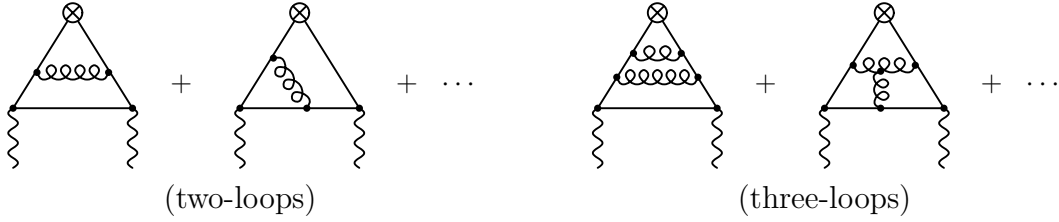

(two-loops)   (three-loops)

Figure 2: Two- and three-loop QCD corrections to the basic triangle diagram.

a tree-graph calculation and we obtain to order $e^2$

$$R^{\sigma\lambda\tau}(-p,p) = \left(-\frac{e^2}{16\pi^2} 4\right) \times [-2i\epsilon^{\sigma\lambda\tau\rho} p_\rho] \ . \tag{4.8}$$

Now we take into account the quark charge factors and the color degrees of freedom and make the following replacement:

$$e^2 \longrightarrow 3\sum_i (e_i e)^2 = 3e^2 n_f \langle e^2 \rangle \qquad \text{for} \quad J_{5S} \ , \tag{4.9}$$

$$e^2 \longrightarrow 3\sum_i \left(e_i^2 - \langle e^2 \rangle\right)(e_i e)^2 = 3e^2 n_f \left(\langle e^4 \rangle - \langle e^2 \rangle^2\right) \quad \text{for} \quad J_{5NS} \ . \tag{4.10}$$

Finally considering the convention to define the photon matrix element as is shown in Eq.(2.2), we reach the expressions (4.1) for $A_S^{(0)}$ and $A_{NS}^{(0)}$.

To obtain the NLO and NNLO terms $A_{NS}^{(1)}$ and $A_{NS}^{(2)}$ corresponding to the flavor nonsinglet axial current $J_{5NS}^\sigma$, we resort to the nonrenormalization theorem [22] for the triangle anomaly. The theorem says that there are no radiative corrections (in this case, QCD corrections) to the basic triangle diagram. The contributions of the two-loop and three-loop diagrams such as shown in Fig. 2 sum up to zero. Moreover, the three-loop diagrams such as shown in Fig. 3, in which the axial current part is connected to the photon-vertex part by two gluon lines, do not contribute to $A_{NS}^{(2)}$. Thus we have

$$A_{NS}^{(1)} = A_{NS}^{(2)} = 0 \ . \tag{4.11}$$

## 4.2 The AB theorem for QCD

In order to evaluate the NLO and NNLO terms $A_S^{(1)}$ and $A_S^{(2)}$ for the photon matrix element of $J_{5S}^\sigma$, we use the AB theorem for the singlet axial current in QCD, which

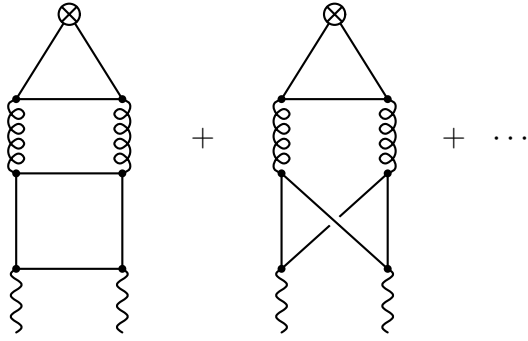

Figure 3: The three-loop diagrams of the gluon-photon scattering type.

reads

$$\partial_\kappa J_{5S}^\kappa = \frac{g^2}{16\pi^2}\frac{n_f}{2}\,\epsilon^{\mu\nu\rho\kappa}G^a_{\mu\nu}G^a_{\rho\kappa}\ , \tag{4.12}$$

where $G^a_{\mu\nu}$ is the gluonic field-strength tensor. This is an operator form and holds to all orders in perturbation theory. We define the following amplitude:

$$\begin{aligned}(p_1+p_2)_\kappa S^{\kappa\lambda\tau}(p_1,p_2) &\equiv\ i\int d^4x_1 d^4x_2 e^{i\{(p_1+p_2)x-p_1x_1-p_2x_2\}}\\ &\quad\times\langle 0|T\Big[\partial_\kappa J_{5S}^\kappa(x)A^\lambda(x_1)A^\tau(x_2)\Big]|0\rangle_{\text{Amputated}}\end{aligned} \tag{4.13}$$

where $A^\lambda$ is a photon field. Differentiating both sides of Eq.(4.13) with respect to $p_{1\sigma}$, and setting $p_1 = -p_2 = -p$, we obtain

$$\begin{aligned}S^{\sigma\lambda\tau}(-p,p) &=\ \int d^4x_1 d^4x_2\ e^{i\ p(x_1-x_2)}\\ &\quad\times\langle 0|T\Big[J_{5S}^\sigma(x)A^\lambda(x_1)A^\tau(x_2)\Big]|0\rangle_{\text{Amputated}}\ ,\end{aligned} \tag{4.14}$$

which is nothing but the photon matrix element of the singlet axial current $J_{5S}^\sigma$ at zero momentum transfer. We consider the amplitude $S^{\sigma\lambda\tau}(-p,p)$ up to the order $e^2 g^4$, namely, up to the three-loop level.

The AB theorem (4.12) tells us that $(p_1+p_2)_\kappa S^{\kappa\lambda\tau}(p_1,p_2)$ is also expressed as

$$\begin{aligned}(p_1+p_2)_\kappa S^{\kappa\lambda\tau}(p_1,p_2) &=\ \frac{g^2}{16\pi^2}\frac{n_f}{2}\,i\int d^4x_1 d^4x_2 e^{i\{(p_1+p_2)x-p_1x_1-p_2x_2\}}\\ &\quad\times\langle 0|T\Big[\epsilon^{\mu\nu\rho\kappa}G^a_{\mu\nu}(x)G^a_{\rho\kappa}(x)A^\lambda(x_1)A^\tau(x_2)\Big]|0\rangle_{\text{Amputated}}\ .\end{aligned} \tag{4.15}$$

The operator $\epsilon^{\mu\nu\rho\kappa}G^a_{\mu\nu}G^a_{\rho\kappa}$ is expanded in terms of gluon fields as

$$\epsilon^{\mu\nu\rho\kappa}G^a_{\mu\nu}G^a_{\rho\kappa} = 4\epsilon^{\mu\nu\rho\kappa}\partial_\mu A^a_\nu \partial_\rho A^a_\kappa + 4gf^{abc}\epsilon^{\mu\nu\rho\kappa}\partial_\mu A^a_\nu A^b_\rho A^c_\kappa ,  \qquad (4.16)$$

where the quartic term of gluon fields does not appear due to the fact that

$$g^2 f^{abc} f^{ade} \epsilon^{\mu\nu\rho\kappa} A^b_\mu A^c_\nu A^d_\rho A^e_\kappa = 0 .$$

Now we find that, up to the order $e^2 g^4$, Eq.(4.15) can be rewritten as

$$\begin{aligned}(p_1+p_2)_\kappa S^{\kappa\lambda\tau}(p_1,p_2) &= \frac{g^2}{16\pi^2}\frac{n_f}{2} \, i \int d^4x_1 d^4x_2 e^{i\{(p_1+p_2)x - p_1 x_1 - p_2 x_2\}} \\ &\times 4\epsilon^{\mu\nu\rho\kappa} \langle 0|T\big[\partial_\mu A^a_\nu(x)\partial_\rho A^a_\kappa(x) A^\lambda(x_1) A^\tau(x_2)\big]|0\rangle_{\text{Amputated}} ,\end{aligned} \qquad (4.17)$$

since photon does not couple to gluon field directly. Again differentiating both sides of Eq.(4.17) with respect to $p_{1\sigma}$, and setting $p_1 = -p_2 = -p$, we obtain up to the order $e^2 g^4$,

$$\begin{aligned}S^{\sigma\lambda\tau}(-p,p) &= \frac{g^2}{4\pi^2}\frac{n_f}{2} \int d^4x_1 d^4x_2 \, e^{i\, p(x_1-x_2)} \\ &\times \epsilon^{\sigma\nu\rho\kappa} \langle 0|T\big[A^a_\nu(x)\partial_\rho A^a_\kappa(x) A^\lambda(x_1) A^\tau(x_2)\big]|0\rangle_{\text{Amputated}} .\end{aligned} \qquad (4.18)$$

The NLO term $A_S^{(1)}$ is derived from the calculation of $S^{\sigma\lambda\tau}(-p,p)$ in the order $e^2 g^2$. Instead of evaluating the r.h.s. of Eq.(4.14), which is a two-loop calculation, we compute the r.h.s. of Eq.(4.18). Then the calculation reduces to the one-loop level but gives null result in this order since gluon and photon do not couple directly. Thus we obtain

$$A_S^{(1)} = 0 . \qquad (4.19)$$

## 4.3 Calculation of $A_S^{(2)}$

In order to obtain $A_S^{(2)}$, we need to calculate $S^{\sigma\lambda\tau}(-p,p)$ in the order $e^2 g^4$. Again instead of evaluating the r.h.s. of Eq.(4.14), which is a three-loop calculation, we compute the r.h.s. of Eq.(4.18). Then the calculation reduces to the two-loop level.

The relevant two-loop diagrams are shown in Fig. 4. There are also two counterterm diagrams corresponding to the diagrams (a) and (b). Because the quarks are

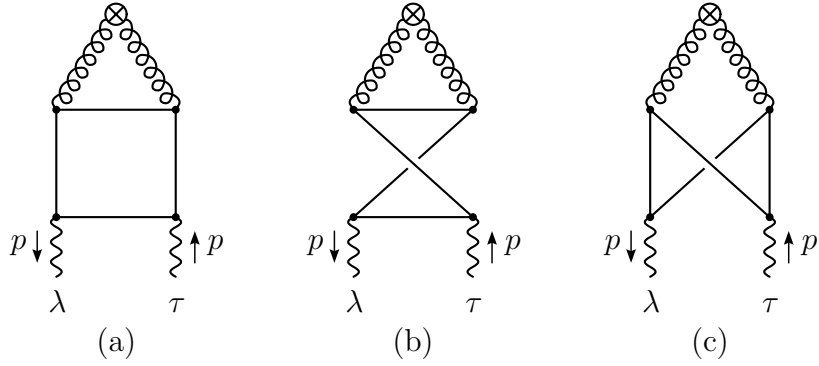

Figure 4: Two-loop diagrams contributing to $S^{\sigma\lambda\tau}(-p,p)$ in the order $e^2 g^4$.

Dirac fermions, there are two directions for the charge flow. The result for each diagram is invariant under reversing the charge flow, so each diagram is computed only for one arrow direction and then multiplied by a factor of two. We have performed calculation in the $\overline{\text{MS}}$ scheme with $n = 4 - 2\epsilon$, and used the latest version of the program FORM [37] to do the necessary algebra.

For the moment let us assume that each quark has a charge $e$. The contributions of the diagrams (a), (b) and (c) are

$$\text{Diagram(a)} = \text{Diagram(b)}$$
$$= \left[\frac{\alpha_s}{\pi}\frac{n_f}{2}\right]\left[\frac{\alpha}{4\pi}\frac{\alpha_s}{4\pi}\right]3C_F S^{2\epsilon}\left(\frac{\mu^2}{-p^2}\right)^{2\epsilon}\left\{\frac{4}{\epsilon} + 16\right\}[-2i\epsilon^{\sigma\lambda\tau\rho}p_\rho] , \quad (4.20)$$

$$\text{Diagram(c)} = \left[\frac{\alpha_s}{\pi}\frac{n_f}{2}\right]\left[\frac{\alpha}{4\pi}\frac{\alpha_s}{4\pi}\right]3C_F S^{2\epsilon}\left(\frac{\mu^2}{-p^2}\right)^{2\epsilon}\left\{\frac{4}{\epsilon} + \frac{82}{3} - 16\zeta_3\right\}[-2i\epsilon^{\sigma\lambda\tau\rho}p_\rho], (4.21)$$

where

$$S^\epsilon = (4\pi)^\epsilon e^{-\epsilon\gamma_E} , \quad (4.22)$$

and a factor $3C_F$ comes from $\text{Tr}[T^a T^a]$ with $T^a$ being the generator of the color group $SU_C(3)$. The contributions of the counter-term diagrams corresponding to the diagrams (a) and (b) are the same and each gives,

$$\left[\frac{\alpha_s}{\pi}\frac{n_f}{2}\right]\left[\frac{\alpha}{4\pi}\frac{\alpha_s}{4\pi}\right]3C_F S^{2\epsilon}\left(\frac{\mu^2}{-p^2}\right)^{\epsilon}\left\{-\frac{6}{\epsilon} - 12\right\}[-2i\epsilon^{\sigma\lambda\tau\rho}p_\rho] . \quad (4.23)$$

Summing up all contributions, we obtain in the order of $e^2 g^4$

$$S^{\sigma\lambda\tau}(-p,p) = \frac{\alpha}{4\pi}\Big(\frac{\alpha_s}{4\pi}\Big)^2 6 C_F n_f \left\{\frac{106}{3} - 16\zeta_3 + 12 \ln\frac{\mu^2}{-p^2}\right\}[-2i\epsilon^{\sigma\lambda\tau\rho} p_\rho] \,. \quad (4.24)$$

We renormalize the photon matrix element $S^{\sigma\lambda\tau}(-p,p)$ at $\mu^2 = -p^2 = P^2$. Also we take into account the quark charge factor and replace $\alpha$ with $\alpha n_f \langle e^2 \rangle$. Then we find

$$A_S^{(2)} = 12\langle e^2 \rangle C_F n_f^2 \left\{\frac{53}{3} - 8\zeta_3\right\} \,. \quad (4.25)$$

Apart from the charge and $SU_C(3)$ group factors, the result (4.25) is consistent with the QED version of the two-loop calculation by Larin [26] for the matrix element of $\epsilon^{\mu\nu\rho\kappa} G^a_{\mu\nu} G^a_{\rho\kappa}$ between two gluon states (see Eq.(22) of Ref.[26]).

# 5 The NNLO $(\alpha\alpha_s^2)$ corrections

Putting the parameters in Eqs.(3.1-3.2), (3.5-3.7), (3.8), (4.1) and (4.25) into Eq.(2.15), we finally obtain the NNLO $(\alpha\alpha_s^2)$ corrections to the first moment of $g_1^\gamma(x,Q^2,P^2)$:

$$\int_0^1 dx g_1^\gamma(x,Q^2,P^2)$$
$$= -\frac{3\alpha}{\pi}\bigg\{\sum_i^{n_f} e_i^4 \left[1 - \frac{\alpha_s(Q^2)}{\pi}\right] - \frac{2}{\beta_0}\Big(\sum_i^{n_f} e_i^2\Big)^2 \left[\frac{\alpha_s(P^2)}{\pi} - \frac{\alpha_s(Q^2)}{\pi}\right]$$
$$+ \frac{2}{\beta_0}\Big(\sum_i^{n_f} e_i^2\Big)^2 \frac{\alpha_s(Q^2)}{\pi}\left[\frac{\alpha_s(P^2)}{\pi} - \frac{\alpha_s(Q^2)}{\pi}\right]$$
$$+ \frac{1}{4\beta_0}\Big(\frac{\beta_1}{\beta_0} - \frac{59}{3} + \frac{2}{9}n_f\Big)\Big(\sum_i^{n_f} e_i^2\Big)^2 \left[\frac{\alpha_s^2(P^2)}{\pi^2} - \frac{\alpha_s^2(Q^2)}{\pi^2}\right]$$
$$+ \frac{2n_f}{\beta_0^2}\Big(\sum_i^{n_f} e_i^2\Big)^2 \left[\frac{\alpha_s(P^2)}{\pi} - \frac{\alpha_s(Q^2)}{\pi}\right]^2$$
$$- \Big(\frac{55}{12} - \frac{1}{3}n_f\Big)\sum_i^{n_f} e_i^4 \frac{\alpha_s^2(Q^2)}{\pi^2} + \Big(\frac{2}{3}\zeta_3 + \frac{1}{36}\Big)\Big(\sum_i^{n_f} e_i^2\Big)^2 \frac{\alpha_s^2(Q^2)}{\pi^2}$$
$$- \frac{1}{12}\Big(\frac{53}{3} - 8\zeta_3\Big)\Big(\sum_i^{n_f} e_i^2\Big)^2 \frac{\alpha_s^2(P^2)}{\pi^2}\bigg\} \,, \quad (5.1)$$

where we are back to the notations $\sum_i^{n_f} e_i^2 \,(= n_f \langle e^2 \rangle)$ and $\sum_i^{n_f} e_i^4 \,(= n_f \langle e^4 \rangle)$. The third to eighth terms in the parentheses are the NNLO contributions. The sum rule

| $n_f$ | $c_0$ | $c_{1Q}$ | $c_{1P}$ | $c_{2Q}$ | $c_{2QP}$ | $c_{2P}$ |
|---|---|---|---|---|---|---|
| 3 | 0.22222 | -0.12346 | -0.098765 | -0.34685 | 0.032922 | -0.41201 |
| 4 | 0.41975 | -0.12346 | -0.29630 | -0.027306 | 0.011852 | -1.1533 |
| 5 | 0.43210 | -0.042405 | -0.38969 | 0.50097 | -0.11860 | -1.4062 |

Table 1: The coefficients in Eq.(5.2) for the first moment of $g_1^\gamma(x, Q^2, P^2)$.

is expressed in the form as

$$\int_0^1 dx g_1^\gamma(x, Q^2, P^2)$$
$$= -\frac{3\alpha}{\pi}\left\{c_0 + c_{1Q}\frac{\alpha_s(Q^2)}{\pi} + c_{1P}\frac{\alpha_s(P^2)}{\pi}\right.$$
$$\left.+ c_{2Q}\left(\frac{\alpha_s(Q^2)}{\pi}\right)^2 + c_{2QP}\frac{\alpha_s(Q^2)}{\pi}\frac{\alpha_s(P^2)}{\pi} + c_{2P}\left(\frac{\alpha_s(P^2)}{\pi}\right)^2\right\}, \quad (5.2)$$

where the coefficients $c_0$, $c_{1Q}$, $c_{1P}$, $c_{2Q}$, $c_{2QP}$ and $c_{2P}$ depend on the number of the active quark flavors, $n_f$. We list in Table 1 the numerical values of the coefficients $c_i$ for $n_f = 3, 4, 5$. We see that the values of $c_i$ increase or decrease with $n_f$. We also see that the values of $c_{2P}$ are rather large in magnitude compared with other coefficients. This is due to the effect of $A_S^{(2)}$, the *three-loop* photon matrix element of the flavor singlet axial current $J_{5S}^\sigma$. Indeed, without the contribution of $A_S^{(2)}$, the numerical values of $c_{2P}$ read as -0.11386, -0.32510 and -0.40405 for $n_f = 3$, 4 and 5, respectively.

To estimate the sizes of the NLO ($\alpha\alpha_s$) and NNLO ($\alpha\alpha_s^2$) corrections compared to the LO ($\alpha$) term and the ratio of the NNLO to the sum of the LO and NLO contributions, we take, for instance, $Q^2 = 30$ and 100 GeV$^2$, and $P^2 = 1$ and 3 GeV$^2$, although these values of $Q^2$ and $P^2$ may not be large enough to apply the formula (5.1) or (5.2) to the cases of $n_f = 4$ and 5. The corresponding values of $\alpha_s$ are obtained from Ref.[23]. We get $\alpha_s(Q^2 = 30\text{GeV}^2) = 0.2048$, $\alpha_s(Q^2 = 100\text{GeV}^2) = 0.1762$, $\alpha_s(P^2 = 1\text{GeV}^2) = 0.4996$, and $\alpha_s(P^2 = 3\text{GeV}^2) = 0.3211$. The results are given in Table 2. Since the numerical values of $c_{2P}$ are rather large in magnitude compared with those of other coefficients, the NNLO corrections become large when $P^2$ is small but still satisfies the condition (1.1). In fact, when $P^2 = 1\text{GeV}^2$ and $Q^2 = 30 \sim 100\text{GeV}^2$, the NNLO corrections amount to 6% ($n_f = 3$), 8% ($n_f = 4$) and $9 \sim 10\%$ ($n_f = 5$) of the sum of the LO and NLO contributions. On the other hand, when $P^2 = 3\text{GeV}^2$, $Q^2 = 30 \sim 100\text{GeV}^2$ and $n_f$ is three to five, the NNLO corrections are found to be about 3% of the sum of the LO and NLO contributions.

|  | $Q^2/\text{GeV}^2$ | $P^2/\text{GeV}^2$ | LO | NLO | NNLO | NNLO/(LO+NLO) |
|---|---|---|---|---|---|---|
| $n_f = 3$ | 30 | 1 | 1 | -0.107 | -0.0520 | -0.0582 |
|  | 100 | 1 | 1 | -0.102 | -0.0505 | -0.0562 |
|  | 30 | 3 | 1 | -0.0816 | -0.0250 | -0.0272 |
|  | 100 | 3 | 1 | -0.0766 | -0.0234 | -0.0254 |
| $n_f = 4$ | 30 | 1 | 1 | -0.131 | -0.0695 | -0.0800 |
|  | 100 | 1 | 1 | -0.129 | -0.0694 | -0.0797 |
|  | 30 | 3 | 1 | -0.0913 | -0.0288 | -0.0317 |
|  | 100 | 3 | 1 | -0.0886 | -0.0287 | -0.0315 |
| $n_f = 5$ | 30 | 1 | 1 | -0.150 | -0.0802 | -0.0944 |
|  | 100 | 1 | 1 | -0.149 | -0.0811 | -0.0953 |
|  | 30 | 3 | 1 | -0.0986 | -0.0309 | -0.0343 |
|  | 100 | 3 | 1 | -0.0977 | -0.0319 | -0.0354 |

Table 2: The NLO and NNLO corrections relative to LO and the ratio of the NNLO to the sum of the LO and NLO contributions for the first moment of $g_1^\gamma(x, Q^2, P^2)$. The values of the QCD running coupling constant we have used are $\alpha_s(Q^2 = 30\text{GeV}^2) = 0.2048$, $\alpha_s(Q^2 = 100\text{GeV}^2) = 0.1762$, $\alpha_s(P^2 = 1\text{GeV}^2) = 0.4996$, and $\alpha_s(P^2 = 3\text{GeV}^2) = 0.3211$.

## 6 Summary

We have investigated the next-to-next-to-leading order $(\alpha \alpha_s^2)$ corrections to the first moment of the polarized virtual photon structure function $g_1^\gamma(x, Q^2, P^2)$ in the kinematical region $Q^2 \gg P^2 \gg \Lambda^2$ in QCD. All the necessary information on the coefficient functions and anomalous dimensions corresponding to the quark axial currents has been already known, except for the three-loop-level photon matrix element (the finite term) of the flavor singlet quark axial current $J_{5S}^\sigma$. Instead of evaluating the relevant three-loop Feynman diagrams, we resort to the Adler-Bardeen theorem for the axial anomaly, Eq.(4.12). Then calculation reduces to the one in the two-loop level. We evaluate in effect the two-loop diagrams for the photon matrix element of the gluon operator, the r.h.s. of Eq.(4.18).

The $\alpha\alpha_s^2$ corrections are found to be about 3% of the sum of the leading ($\alpha$) and the next-to-leading ($\alpha\alpha_s$) contributions, when $Q^2 = 30 \sim 100 \text{GeV}^2$ and $P^2 = 3\text{GeV}^2$, and the number of active quark flavors $n_f$ is three to five.

## Acknowledgments

We thank O. Tarasov and K. Chetyrkin for valuable discussions. This work is supported in part by Grant-in-Aid for Scientific Research from the Ministry of Education, Culture, Sports, Science and Technology, Japan No.15540266.